\documentclass[aps,twocolumn,floats,floatfix,showpacs,superscriptaddress]{revtex4}
\usepackage{epsfig}
\begin{document}
\title{Assortative model for social networks}
\author{Michele Catanzaro}
\affiliation{INFM UdR ROMA1 Dipartimento di Fisica, Universit\`a di Roma
``La Sapienza'' Piazzale A. Moro 2 00185 Roma, Italy}
\author{Guido Caldarelli}  
\affiliation{INFM UdR ROMA1 Dipartimento di Fisica, Universit\`a di Roma
``La Sapienza'' Piazzale A. Moro 2 00185 Roma, Italy}
\author{Luciano Pietronero} 
\affiliation{INFM UdR ROMA1 Dipartimento di Fisica, Universit\`a di Roma
``La Sapienza'' Piazzale A. Moro 2 00185 Roma, Italy}
\affiliation{CNR Istituto di Acustica, ``O.M. Corbino'', via Fosso del 
Cavaliere 100, 00133 Roma, Italy}
\date{\today} 
\begin{abstract} 
In this paper we present a new version of a network growth model,
generalized in order to describe the behavior of social networks. 
The case of study considered is the preprint archive at 
cul.arxiv.org. 
Each node corresponds to a scientist,
and a link is present whenever two authors wrote a paper together.
This graph is a nice example of degree-assortative network,
that is to say a network where sites with similar degree are 
connected each other. The model presented is one of the few able 
to reproduce such behavior, giving some insight on the microscopic 
dynamics at the basis of the graph structure.
\end{abstract} 
\maketitle

Networks \cite{B85,re3} are present in different phenomena.
The Internet \cite{FFF99,CMP00} is a graph composed by different computers,
connected by cables; the WWW \cite{HA99,BAJ99} is a graph composed by HTML
documents connected by hyperlinks, even social structures \cite{N01,ASBS00}
can be described as graphs. In the latter case the nodes are individuals 
connected by different relationships. Even if the degree probability 
distribution $P(k)$ (i.e. the frequency to find a number $k$ 
of links per node) is very often scale-free (i.e.  $P(k) \propto 
k^{-\gamma}$), other quantities allow to distinguish between the various 
cases. For such purpose, one of the most interesting is the assortativity by degree.
Assortativity can be defined as the tendency for nodes in a social network to
form connections preferentially to others similar to them\cite{Newman}. 
This mechanism has been proposed as the key ingredient for the 
formation of communities in networks\cite{Vespignani,NEW1}. 
Using this quantity, it is possible to distinguish the technological networks, where instead, 
the behavior is rather degree-disassortative, so that 
vertices tend to be linked to others different from them.
Despite the relative simplicity of such behavior few models\cite{BB01,CCDM02,Calla} 
of network growth are able to reproduce the formation of communities 
and no one explains the difference between social and technological networks.

Here we analyze a specific case of social network, namely the ArXiv:cond-mat 
repository of preprints at cul.arxiv.gov collected by Mark Newman\cite{N01}. The nodes are 
the authors of the various papers and a link is present between them whenever  
they wrote at least one paper together. We are able to reproduce most of the 
features of such network by a suitable modification of a model presented in 
Ref.\cite{GUIDO}.
The quantities we measured in the real data and in the model are 
the {\em degree} probability distribution, the {\em degree correlation} 
between neighbor sites, the 
{\em clustering} and the {\em site betweenness} probability distribution. 
A summary of the results is reported in Tab.1.

The degree is the number of links per node. As expected, the degree probability 
distribution of the cond-mat data show a power law behavior of the kind 
$P(k) \propto k^{-\gamma}$ with $\gamma=3$ (see diamonds in Fig.1). 

We then measure the degree correlation between nodes. This is done by introducing 
the quantity $Knn(k)$, giving the average
degree of the site neighbors of one site whose degree is $k$.
$Knn$ increases if nodes are correlated by degree (assortative 
networks). It decreases if they are anti-correlated (disassortative 
networks). It is flat if they are uncorrelated (for example, in the BA model\cite{Annd}).
$Knn$ in the data has an increasing trend, consistent with 
our expectation for an assortative network. 
A power law seems to be an appropriate fit in 
the region of growth $Knn(k) \propto k^{\phi}$ 
where $\phi$ is about $0.2$ (See diamonds in Fig. 2). 
Another measure of assortativity we considered is the 
assortativity coefficient $r$. A complete definition of this 
quantity can be found
in ref.\cite{NEW3}, here we can say that it is proportional to
the connected degree-degree correlation function.
In this paper we find that both $r$ and $\phi$ have the 
same behaviour by varying the parameters of the model. 
We therefore focus our analysis only on the $\phi$.

Clustering coefficient $c_i$ for every site $i$ gives the probability that 
two nearest neighbors of vertex $i$ are also
neighbors each other.
$cc(k)$, is the average clustering coefficient for sites whose degree is $k$, 
and it measures the tendency to form cliques where each nearest 
neighbor of a node (with degree $k$) is connected to each other. 
In real networks this usually decreases with a power-law 
$cc(k) \propto k^{\psi}$ 
($\psi=-0.8$ for the data we analyzed) because hubs tend to play the role of 
connections between separate clusters in the 
graph, i.e. clusters that have few other interconnections than the ones 
passing through the hub. Then the high degree node tends to have low 
clustering coefficient. 

The betweenness $b_i$ of a vertex $i$ gives the probability that the 
site $i$ is in the path between two other vertices in the graph.
Therefore it might be interpreted as
the amount of the role played by the vertex $i$ in social relation between
two persons $j$ and $k$. 
This quantity behaves as a power law both in its distribution 
$P(b) \propto b^{-\eta}$ ($\eta=2.2$) and in 
dependence upon $k$. Analogously to the clustering case we defined the
average betweenness $b(k)$ for vertices whose degree is $k$. From Fig. 3
we find $b(k) \propto k^{\varepsilon}$ with $\varepsilon=1.81$. 

The model we defined in order to reproduce the data is inspired to the 
preferential attachment one\cite{BAJ99}. 
The main variation consists in allowing growth by addition of new links 
between old nodes. More particularly at every step of growth:
\begin{enumerate}
\item{with probability $p$ a new node is wired to an existing one; 
the choice of the destination node is left to Barab\'asi-Albert preferential
attachment rule('rich gets richer').
Thus the probability of adding a new node and connecting it to an old node $i$
is}
\begin{equation}
p\frac{k_i}{\sum_{j=1,N} k_j}.
\end{equation}
\item{with probability $(1-p)$ a new edge is added (if absent) between two existing 
nodes.
These are chosen on the basis of their degree. In other words,
the probability of adding an edge between node $1$ and node $2$ is a $\tilde{P}(k_1,k_2)$.
This can be written as $P_1(k_1)P_2(k_2|k_1)$, being the second factor a 
conditioned probability. 
$P_1(k_1)$ is the rule for choosing the first of the two nodes, and again
it is determined by the preferential attachment. The functional form of 
$P_2(k_2|k_1)$ can be chosen so as to favor links between similar or
different degree.
In this way, the probability of adding a new edge and connecting two old
non-linked nodes is} 
\begin{equation}
(1-p)\frac{k_i}{\sum_{j=1,N} k_j}P_2(k_2|k_1)
\end{equation}
\end{enumerate} 

In the limit of $p=1$ the model reduces to a traditional BA tree.
In order to reproduce the assortative behavior we have explored two different functional forms:
an inverse dependence 
\begin{equation}
P_2(k_2|k_1)\propto \frac {1}{|k_1-k_2|+1}
\end{equation}
and an exponential dependence, which clearly has a stronger effect
\begin{equation}
P_2(k_2|k_1)\propto e^{-|k_1-k_2|}.
\end{equation}

Results of simulations for the various values of $p$ are summarized 
in Tab.1, where  the 
fitted exponents of the distributions and the global quantities 
describing the networks are reported.
As $p$ grows from $0.1$ to $1.0$ the change in the statistical properties 
is consistent with the rough estimate for 
the degree distribution exponent given in Ref.\cite{GUIDO} 

\begin{equation}
\gamma(p) = 2+\frac{p}{2-p}
\end{equation}

As $p$ tends to 1.0, the exponent approaches the value 3 of the BA model. 
A radically different behavior appears in the exponential case. 
While for high $p$ we still have scale-free distribution, as $p$ decreases
a structure in $k$ emerges. Two regimes become visible: 
a power-law distribution for low $k$ and a peaked distribution for high $k$.

Similar behavior is evident for all the quantities depending on $k$.
The transition happens around $p=0.5$. 
This behavior can be explained as follows. Edges are added
mainly between high degree nodes because of the 'preferential
attachment option' adopted in the choice of the first vertex. 
Moreover, the strong assortativity deriving from the exponential form 
imposes  an high degree to the second node as well.
Therefore, when the 'wiring component' of the growth prevails ($p$ below $0.5$), 
a cluster of hubs appears. 
Their degrees are sharply distributed around a high value.   
Thus a strong assortativity can break up the self-similar structure of 
the graph, superimposing a distribution with a typical scale on the 
scale-free one. This highlights the typical aspect of an assortative network, 
where  
the hubs (highly connected nodes) connect with other hubs, generating a 
core-periphery structure. This structure is emphasized 
in the exponential case, where assortativity becomes so large to induce a phase
transition from a scale-free graph to a network with a characteristic scale for 
high degrees. 

The slope of $Knn(k)$ grows as the assortativity is increased, moving from the 
inverse to the exponential form, and reducing the value of $p$.
The slight inversion in the growth of the exponent visible at small $p$
can be explained as a finite size effect, highlighted by the intense 
assortativity for very low values of the parameter $p$.
The BA limit is visible as well, being the distribution roughly flat for 
$p=1.0$.
By measuring $\phi$ and $r$ we note that their trends, as
the parameters change, are analogous. Reasonably enough, we can conclude
that, at least for our model, the exponent and the coefficient carry the same
information.

The clustering coefficient distribution versus the degree fails to reproduce 
the real trends. These are usually decreasing with a power-law; the model,
instead, generates increasing trends. We fit them with a power law
with positive exponent.
We can explain qualitatively such incongruence by taking into account high degree vertices.
In real networks hubs tend to play the role of connections between separate 
clusters in the graph, with few links between each other
(apart from the ones attached to the hub). Therefore this nodes tend to have 
low clustering coefficient. 
In our model, on the other hand, all the hubs are aggregated 
together. Thus, even producing an assortative network it cannot reproduce 
a network with $cc(k)$ decreasing with $k$.
We comment that such behavior in the real data is due to the different areas 
of expertise of various authors, such that the most productive scientists in 
one discipline do not collaborate with the top scientists of other disciplines 
within cond-mat. 
Imposing such separation on the hubs produced by the model 
reproduces the correct behavior of data (or rather analyzing the data by dividing 
the papers according to the fields).

As regards the betweenness, $b(k)$ is an increasing function of
$k$ (hubs are crucial in the exchange of information). 
On the other hand
its slope decreases as p is reduced. In a tree like structure ($p=1.0$),
hubs are play the role of bottlenecks for the flow of information 
between separate parts of the networks. 
Therefore, they have very high site betweenness. 
Approaching to a core-periphery structure, each node of the core becomes 
approximately as good as the others in performing this job. Therefore the 
site betweenness of high degree nodes decreases. 

The site betweenness distribution $P(b)$ or is plotted after integration 
in Fig.3. 
We obtain a power-law with an exponent not depending significantly on $p$.
Its averaged value is 2.0, that is equal to the measured value for a BA tree
\cite{goh}. 
It is interesting to notice that also here
a characteristic scale appears at high values of the site betweenness.
This is visible in the bump 
that distorts the scale
free nature of the integrated distribution.
Notice that we would see a similar distorted trend if we integrated
the degree distribution. 

In ref.\cite{goh2} the following scaling relation is demonstrated for the BA
model
\begin{equation}
b \propto k^{(\gamma-1)/(\eta-1)}
\end{equation}
Thus, the exponent of the site betweenness plotted versus $k$ is related
to the previous two by the equality
\begin{equation}
\varepsilon=(\gamma-1)/(\eta-1)
\end{equation}
This relation stands for disassortative and not assortative 
networks, while deviations are shown for assortative ones in ref.\cite{goh3}.
By computing this difference we noticed a slightly growing trend, 
as $p$ is decreased, 
giving further evidence that assortativity breaks the scaling relation.

The qualitative agreement between the distribution of the real data and 
the simulation shows that our model is able to catch the 
basic aspects of the real graph, with the only above mentioned 
exception of the clustering coefficient versus $k$. 
A quantitative comparison suggests that the exponential
form is too strong to describe existing networks.
In fact, the appearance of a characteristic-scale structure like the one 
foreseen in our model has not been observed in any of the real assortative
networks studied until now. One must notice as well the slight difference
in the exponents of the site betweenness distribution (2.0 for the simulation
and 2.2 for cond-mat). Following ref.\cite{goh},  
networks should be divided in two classes of universality according to the 
exponent of their site betweenness distribution.
In fact this seems to assume always one of the two values 2.0 and 2.2.
Co-authorship networks fall in the second class. Therefore, 
if the hypothesis of ref.\cite{goh} were confirmed, our model would fail guess
the correct universality class for the networks that it is
thought to represent. However, this would be reasonable, since
the model can be reduced to a BA tree, which falls in the first class.

In conclusion, we have studied a generalized graph growth model,
where by tuning a parameter $p$, it is possible to weight the role of
growing (addition of new nodes) and mixing (addition of new edges) in the
microscopical behavior of the network. The assortativity can be 
controlled as well by fixing a functional form for the
wiring probability. Macroscopic characteristics of the network,
i.e. statistical distributions, have been derived by simulations in the
assortative case.
The results reveal the effects of assortativity on the topology of 
a network, that can be as dramatical as a phase transition.
Moreover, the simulation succeed in reproducing most of the features of real assortative networks.
Future work could focus on many aspects:
new nodes could be added carrying $2$ edges instead of one, in order to have 
a BA graph rather than a BA tree in the $p=1.0$ limit;
the rate of addition of new nodes and of new links could be 
measured for real networks to have a fine tuning of the parameter $p$;
more general functional forms for the wiring could be investigated,
and even the preferential attachment choice could be changed,
in order to have a significant wiring also for low degree nodes. 
Further extensions are possible because of the rich flexibility of the
model.

We thank the FET Open project IST-2001-33555 COSIN.

\begin{figure}[htbp]
\centerline{\psfig{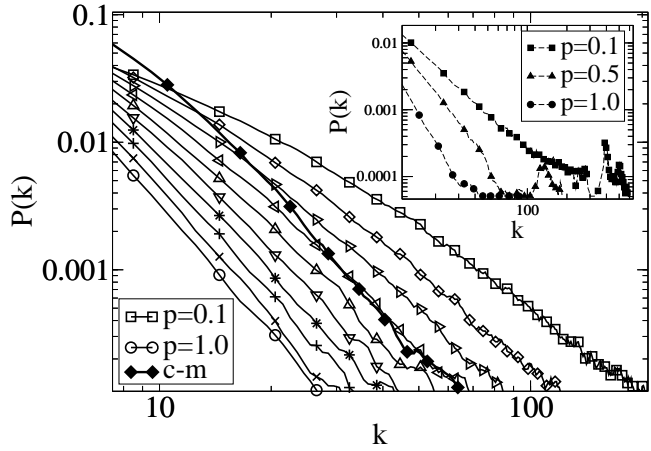}} 
\caption{Degree distribution in the inverse case. The slope increases momotonically
as p grows from 0.1 to 1.0. The distribution for cond-mat is reported for 
comparison. In the inset, degree distribution in the 
exponential case. As p becomes smaller than 0.5 a peaked structure at high 
degrees appears.}
\label{fig3} 
\end{figure}

\begin{figure}[htbp] 
\centerline{\psfig{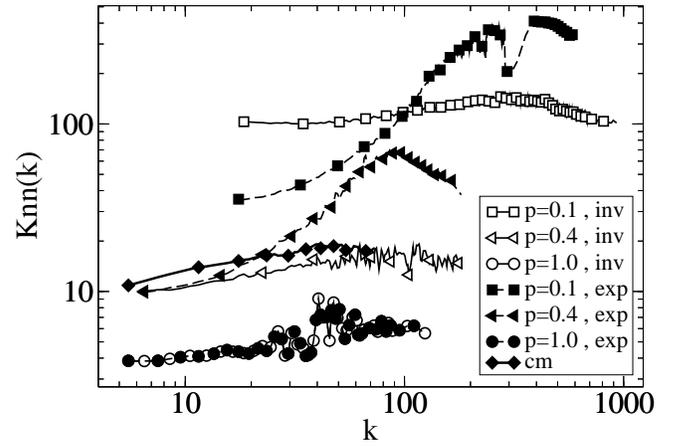}} 
\caption{Average nearest neighbour degree versus k in the inverse and 
exponential case, and for cond-mat. In the exponential case a structure at 
high $k$ is visible for low $p$. For cond-mat distribution, a maximal and 
a minimal slope can be defined.}
\label{fig4} 
\end{figure}

\begin{figure}[htbp] 
\centerline{\psfig{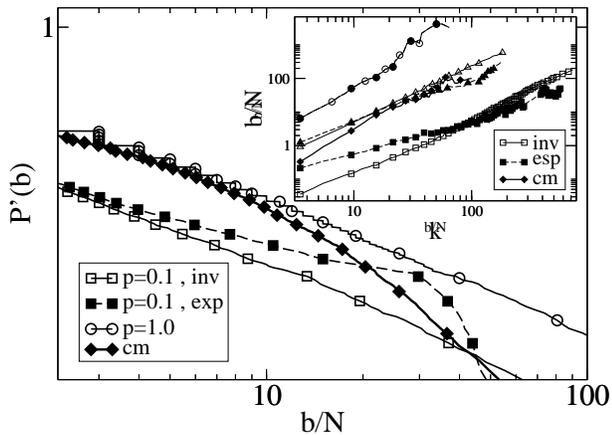}} 
\caption{Integrated site betweenness distribution in the inverse and 
exponential case, and for cond-mat. 
As $p$ tends to $1.0$ the branching in the graphs increases. Given a branch of 
$n$ nodes, $b_n$ starting from the leaves is proportional 
to $(N-1)$, $2(N-2)$, $4(N-3)$,..$2^{(n-1)}(N-n)$.
Consequently, in a tree-like structure the site betweenness is 
quantized. This appears in the distribution as 
a succession of power law distributed spikes (stairs in the integrated distribution). 
For small p, a bump is visible, signalling a characteristic scale.
In the inset, $b$ versus $k$ in the inverse and 
exponential case, and for cond-mat. In the exponential case a structure at 
high $k$ is visible for low $p$.}
\label{fig5} 
\end{figure}

\begin{table}[h]
\begin{centering}
\caption {Results of numerical simulation of the model: exponents of the 
distributions and assortativity coefficient. Last row refers to cond-mat co-authorship network. The 
exponent of the site betweenness distribution is not reported since its 
fluctuations around the average value of 2.0 are negligible. For cond-mat 
it is 2.2. $\rho=2+\frac{p}{2-p}$ and 
$\mu=|\varepsilon-\frac{\gamma-1}{\eta-1}|$
The error on the figures is always less than $5\%$.}
\label{tab1}
\begin{tabular}{|*{12}{c|}}
\hline
$p$        &$\rho$  &$\gamma_{inv}$  &$\gamma_{esp}$  &$\phi_{inv}$  &$\phi_{esp}$  &$\psi_{inv}$  &$\psi_{esp}$  &$\varepsilon_{inv}$  &$\varepsilon_{esp}$  &$\mu_{inv}$  &$\mu_{esp}$ \\
\hline
$0.1$      &$2.05$             &$2.05$          &$1.73$          &$0.23$        &$0.90$        &$0.58$        &$2.31$        &$1.71$               &$0.94$               &$0.62$       &$0.21$ \\ 
$0.2$      &$2.11$             &$2.27$          &$1.83$          &$0.24$        &$0.87$        &$0.61$        &$2.47$        &$1.65$               &$1.09$               &$0.38$       &$0.26$ \\ 
$0.3$      &$2.18$             &$2.33$          &$2.18$          &$0.25$        &$0.88$        &$0.65$        &$2.69$        &$1.63$               &$1.16$               &$0.30$       &$0.02$ \\ 
$0.4$      &$2.25$             &$2.52$          &$2.33$          &$0.25$        &$0.89$        &$0.73$        &$2.78$        &$1.64$               &$1.27$               &$0.12$       &$0.06$ \\ 
$0.5$      &$2.33$             &$2.61$          &$2.45$          &$0.25$        &$0.90$        &$0.67$        &$2.97$        &$1.66$               &$1.34$               &$-$          &$0.11$ \\ 
$0.6$      &$2.43$             &$2.78$          &$2.59$          &$0.23$        &$0.85$        &$0.81$        &$2.90$        &$1.70$               &$1.50$               &$-$          &$-$    \\ 
$0.7$      &$2.54$             &$2.87$          &$2.71$          &$0.23$        &$0.84$        &$0.74$        &$3.10$        &$1.73$               &$1.61$               &$-$          &$-$    \\ 
$0.8$      &$2.67$             &$2.92$          &$2.83$          &$0.21$        &$0.76$        &$-$           &$3.50$        &$1.77$               &$1.71$               &$-$          &$-$    \\ 
$0.9$      &$2.82$             &$2.96$          &$2.94$          &$0.16$        &$0.67$        &$-$           &$-$           &$1.84$               &$1.88$               &$-$          &$-$    \\ 
$1.0$      &$3.00$             &$3.01$          &$3.09$          &$0$           &$0$           &$-$           &$-$           &$2.06$               &$1.99$               &$-$           &$-$   \\ 
\hline
cm 	   &$-$                &\multicolumn{2}{c|}{2.99}        &\multicolumn{2}{c|}{0.14-0.35}&\multicolumn{2}{c|}{-0.80}   &\multicolumn{2}{c|}{1.81}                  &\multicolumn{2}{c|}{0.41} \\
\hline
\end{tabular}
\end{centering}
\end{table} 

\end{document}